\documentclass[12pt]{article} 
\title{Gravity, Cosmic Rays and the LHC}  
\author{{\it Richard Shurtleff~}\thanks{affiliation and mailing 
address: Department of Applied Mathematics and Sciences, 
Wentworth Institute of Technology, 550 Huntington Avenue, 
Boston, MA, USA, ZIP 02115, telephone number: (617) 989-4338, e-mail address: shurtleffr@wit.edu}} 
\usepackage{graphicx}
\graphicspath{{converted_graphics/}}
\begin{document} 
          
\maketitle 

\begin{abstract} 
The high energy proton beams expected when the Large Hadron Collider (LHC) comes online should provide a pass/fail test for a gravity-related explanation of ultrahigh energy cosmic rays. The model predicts that particles have two kinds energies, equal for null gravitational potentials and, in the potential at the Earth, differing significantly above one TeV. If correct, a 7 TeV trajectory energy proton at the LHC would deliver a 23.5 TeV particle state energy in a collision. 

PACS -  11.10.-z, 29.20.D-, 96.50.S

Keywords -  field theory, proton accelerators, Large Hadron Collider, cosmic rays

\end{abstract}

As proton beams in accelerators achieve energies previously the exclusive domain of cosmic rays (CRs), experiments in the lab can begin to sort out the various explanations of what many consider to be the mysterious behavior of ultrahigh energy CRs. The 7 TeV proton beam expected shortly at the Large Hadron Collider crosses a gravitational threshold that should confirm or discard at least one CR explanation. 

That explanation is presented in Ref. \cite{I}, hereafter referred to as `I'. To give some idea of what it is about, observe that the effects of electromagnetic fields on protons in a particle accelerator and in interstellar space produce nearly classical trajectories. And a classical trajectory in quantum mechanics maximizes or minimizes the phase of the quantum field. The term `trajectory (four-) momentum' refers to the spatial momentum and energy of the proton moving along its trajectory.

Now consider a second kind of momentum. Also from quantum mechanics we know that particle states can be expanded over a basis of momentum eigenstates. Each eigenstate is proportional to a plane wave whose phase is the scalar product of momentum $p$ and Minkowski event coordinates $x$, $\exp{(ip\cdot x)}.$ Then momentum is the rate of change of phase with respect to distance and time, which we call the `particle state (four-) momentum'.

In the quantum field theory of free massive particles the phases of the quantum field that determine the trajectory and the phases of the particle states match so that the trajectory momentum and the particle state momenta correspond. In I and a previous work \cite{FFF}, a well-known construction of quantum fields in flat spacetime \cite{W} is modified by including more general translation representations. This should be allowed since translations along with rotations and boosts form the group of spacetime symmetries connected to the identity. The separation of particle state and trajectory momenta can be traced back to the fact that {\it{any translation preserves all coordinate differences.}}

The modification introduces additional free parameters. Then assumptions are made in I so that the trajectory momentum and particle state momentum can explain the cosmic ray spectrum by having one energy greater than the other in a gravitational field.

The time component of momentum is the total energy. For the high energy particles considered here, the total energy is much larger than the rest energy and the kinetic energy is just a little less than the total energy. So we can refer to either total or kinetic energy as simply the `energy'. 

In I, the particle state energy $\bar{E}$ depends on the trajectory energy $E$ and the gravitational potential $\phi.$ To the accuracy needed here, one has from I the relation
\begin{equation} \label{energy}
\bar{E} = E(1 - 4 \phi \gamma^{2}) \, ,
\end{equation}
where the gravitational potential includes a factor of the square of the speed of light $c$ to make the quantity unitless and $\gamma$ is the relativistic gamma, $\gamma$ =  $1 + E/mc^{2} \approx$ $E/mc^{2},$ since the rest energy $mc^{2}$ is small compared to $E.$ The value of the gravitational potential at the Earth's surface is $-1.06 \times 10^{-8},$ referenced to a null potential in interstellar space. Because $\mid \phi \mid$ is so small, terms of higher order in $\phi$ have been dropped.

To apply Eq. (\ref{energy}) to a CR proton primary, it is argued that the trajectory energy $E$ is the proton's energy upon acceleration most likely in interstellar space at a supernova remnant's shock wave.\cite{xrays,Hillas} The proton then travels in the disk of the Galaxy with this same trajectory energy $E,$ trapped by the galactic magnetic field. As it approaches the Earth, the trajectory energy $E$ increases by a negligible amount due to the conventional effects of gravity. Thus the CR proton strikes the Earth's atmosphere with a trajectory energy $E,$ essentially unchanged since its acceleration at the supernova remnant.

In interstellar space, the particle state energy $\bar{E}$ is the same as the trajectory energy $E$ because $\phi$ vanishes in interstellar space, i.e. $\bar{E}$ = $E$ when $\phi$ = 0. As the CR proton approaches the Sun and Earth, the particle state energy begins to increase because the potential $\phi$ drops below zero. When the proton interacts with the atmosphere it delivers the energy $\bar{E}$ creating a CR shower. For a trajectory energy of $E$ = 2 PeV = $2 \times 10^{15}$ eV, the proton deposits $\bar{E}$ = 300 EeV = $ 3 \times 10^{20}$ eV of particle state energy as can be verified with Eq. (\ref{energy}). The trajectory energy 2 PeV is near the limit of expected energies from supernova remnant accelerations, and the 300 EeV energy is the energy of the most energetic CR particle so far detected. \cite{OMG} Thus an ultrahigh energy CR proton primary is in an ultrahigh energy particle state only when  near the Sun and Earth. 

One indication that the model in I is wrong is the observed anisotropy of incident CRs. The Pierre Auger Collaboration recently determined that ultrahigh energy CRs seem to arrive from Active Galactic Nuclei or other extragalactic sources. \cite{Auger} 

However, a stronger test of the model is expected when the Large Hadron Collider (LHC) \cite{LHC} becomes operational. See Fig. 1 for a graph of the energy ratio $\bar{E}/E$ from Eq. (\ref{energy}) in the upper energy region of currently operating or proposed proton accelerators. 

For an $E$ = 7 TeV proton trajectory energy at the LHC, Eq. (\ref{energy}) predicts a proton particle state energy of 23.5 TeV. The proton bunches move, as designed, through the magnets and other electromagnetic fields with nominal 7 TeV trajectories. But when they interact with other protons or a target, the particle state energy applies and they deliver 23.5 TeV per proton to the target. By comparison, an $E$ = 1 TeV trajectory proton at the Tevatron \cite{Tevatron} would gain only 5\% and deliver 1.05 TeV per proton, not much more than expected. The predicted factor of more than 3 in beam energy from 7 to 23.5 TeV at the LHC should make it obvious whether or not the assumptions made in I could be valid.

\begin{figure}[tbp] 
  \centering
  \includegraphics[bb=0 0 360 253,width=5.67in,height=3.98in,keepaspectratio]{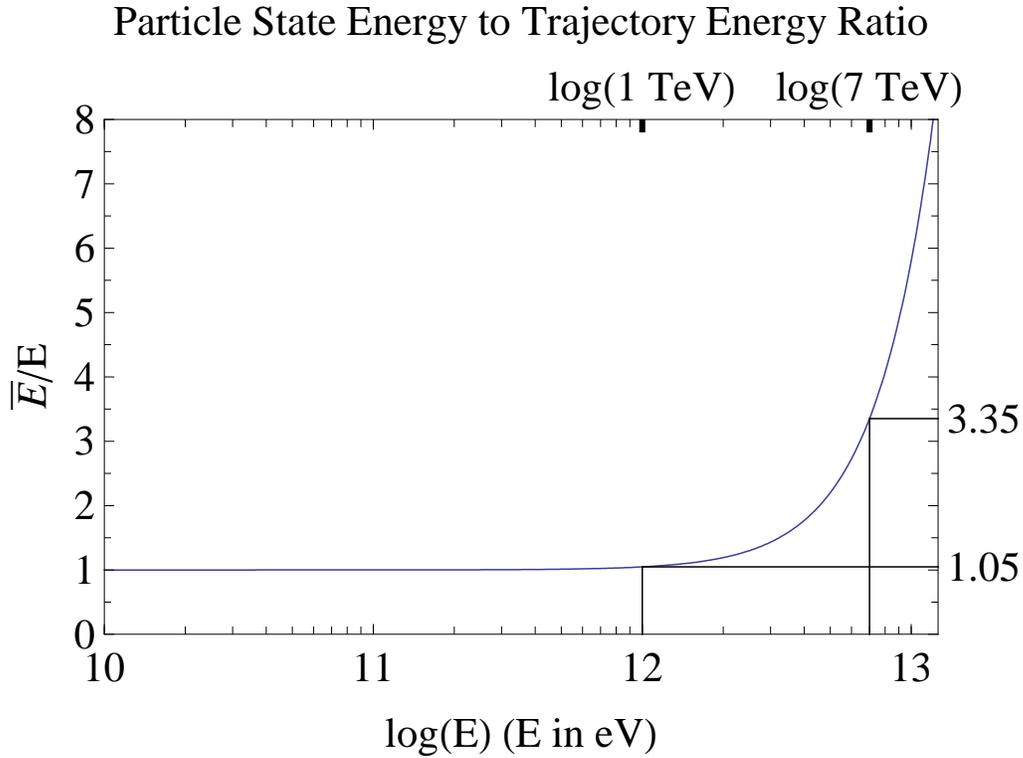}
  \caption{The Two Energies for a Proton. A proton or other massive particle, in Ref. \cite{I}, `I', has two energies whose ratio is $\bar{E}/E$ =  $1-4 \phi \gamma^{2},$ where $\phi$ is the gravitational potential divided by the square of the speed of light and $\gamma$ is the relativistic gamma of the proton's trajectory.  The trajectory energy $E$ and the particle state energy $\bar{E}$ are nearly equal up to about 1 TeV. Doubling is often used as a threshold; here doubling occurs at $E$ = 4.6 TeV giving $\bar{E}$ = 9.2 TeV. Given 7 TeV as a possible proton energy at the LHC, collisions should deliver a particle state energy per proton of 23.5 TeV (= $3.35 \times 7$ ). For protons at the highest energy observed for cosmic rays, the ratio is on the order of $10^{5}.$}
  \label{fig:fig1b}
\end{figure}

\appendix

\section{Problems} \label{Pb}

\vspace{0.3cm}
\noindent 1. According to the LHC website \cite{LHC}, a proton beam can have 2808~bunches with $1.15 \times 10^{11}$ protons each. Find the beam energy in megajoules and liters of gasoline if each proton has (a) 7 TeV of energy and (b) 23.5 TeV of energy.

\vspace{0.3cm}
\noindent 2. To the level of approximation in Eq. (\ref{energy}), the spatial momentum has the same ratio of particle state to trajectory values. Thus $\bar{p}^{\mu}$ = $p^{\mu}(1-4\phi \gamma^{2}),$ where the index runs over $\mu \in$ $\{1,2,3,4\}$ = $\{x,y,z,t\}$ in a Minkowski coordinate system. Find the particle state mass $\bar{m}$ for a proton with a trajectory energy of 7 TeV. [The particle states in I are `effective particle states' derived from the proton particle states.]

\vspace{0.3cm}
\noindent 3. By searching experimental results, find the flux of cosmic rays incident on the Earth's atmosphere that have an energy of 23.5 TeV. Give the answer in units of the number of particles per square meter per steradian per TeV per day. 

\end{document}